# Ultrasmall InGa(As)P dielectric and plasmonic nanolasers

Debarghya Sarkar[1,2,†], Sangyeon Cho[1,2,†], Hao Yan[1,2], Nicola Martino[1,2], Paul H. Dannenberg[1-3], Seok-Hyun Yun[1-3*]

†Equal Contribution
*Corresponding Author: syun@hms.harvard.edu

[1]Harvard Medical School, Boston, MA 02115, USA
[2]Wellman Center for Photomedicine, Massachusetts General Hospital, Boston, MA 02114, USA
[3]Harvard-MIT Health Sciences and Technology, Massachusetts Institute of Technology, Cambridge, MA 02139, USA

**Abstract**

Nanolasers have great potential as both on-chip light sources and optical barcoding particles. We demonstrate ultrasmall InGaP and InGaAsP disk lasers with diameters down to 360 nm (198 nm in height) in the red spectral range. Optically pumped, room-temperature, single-mode lasing was achieved from both disk-on-pillar and isolated particles. When isolated disks were placed on gold, plasmon polariton lasing was obtained with Purcell-enhanced stimulated emission. UV lithography and plasma ashing enabled the fabrication of nanodisks on a wafer-scale, with intended random size variation. Silica-coated nanodisk particles generated stable sub-nanometer spectra from within biological cells across an 80 nm bandwidth from 635 to 715 nm.

**Introduction**

Submicron- and nano-lasers have received interests for basic research and applications including optical communications, integrated photonics, chemical sensing, and biological probing.[1,2] A small device volume is the defining characteristic of nanolasers. The small size not only provides some unique physical phenomena, such as large Purcell effects[3] and threshold-less lasing,[4] but it is also important for practical applications. A small volume of the gain medium is beneficial for low pump energy consumption. Also importantly, smaller sizes are generally preferred for biological applications to facilitate cellular uptake or attachment on the cell membrane with minimal physical perturbations on cells.[5] Furthermore, a small size is critical for the transport and diffusion of nanolasers in tissues.[6,7]

Recently, free-standing miniature lasers, or laser particles (LPs), have emerged as promising new optical probes for imaging, sensing, and mass-scale cellular tagging.[6,8–10] The spectrally narrowband stimulated emission from LPs can constitute several hundreds of frequency (color) channels, providing much enhanced multiplexing capability than conventional fluorescent probes. The multi-color feature makes LPs uniquely suited for cell-tracking analysis.[11] III-V compound semiconductors are an attractive choice of materials for miniature lasers due to their high refractive index and high optical gain,[12] as well as good material stability in aqueous media when coated with a protective silica layer.[10]

Nonetheless, semiconductor nanolasers reported to date have shown two limitations. First, the smallest size of a room-temperature laser has been larger than the optical wavelength in air in at least one dimension, to provide net intracavity gain.[1] The exceptions are resonant Mie scattering nanoparticles made of $CsPbBr_3$ perovskite crystals[13] and spaser (surface plasmon amplification by stimulated emission of radiation) particles involving microbubble cavities.[14] However, the water-solubility of the perovskites and the excessive heating of the spasers make them not suited for nonperturbative biological applications. Second, the fabrication of the nanostructures has almost exclusively required electron beam lithography for III-V semiconductors[6,15], or chemical growth of individual crystals such as for GaN nanowires[16,17] and II-VI CdS/CdSe nanowire[18] or plates[19,20]. The electron beam exposure process is time-consuming and expensive, and thus is not acceptable for high-volume production of LPs for barcoding applications that require a large number of LPs.[10,11] Crystal growth has the potential to meet the production-scale requirements but is less established than nanofabrication, especially to directly create III-V semiconductor nanostructures that are suited for biological samples.

Here, we report significant advances in the fabrication and miniaturization of semiconductor nanolasers. First, we demonstrate wafer scale fabrication of circular nanodisks made of InGa(As)P semiconductor using UV stepper lithography and plasma ashing method. Our scalable fabrication meets the need to produce a large number of nanodisks in the wafer scale with different colors for tagging biological cells. Second, we achieve room temperature lasing from subwavelength nanolasers-on-pillar with diameters as small as 460 nm (198 nm height) at optical wavelengths around 660 nm. This represents about 40% reduction in size compared to previous InGaP quantum-well disks fabricated by electron-beam lithography.[6,15] We further push the size reduction by the help of a gold substrate *via* surface plasmon polariton[20,21] and demonstrate lasing from disks with a diameter of 360 nm. Using a combination of two semiconductor gain materials and diameter variation, we demonstrate lasing from a large number of nanodisk LPs over an 80 nm spectral bandwidth. The multi-color nanodisk LPs coated with biocompatible silica shell and surface functionalized with polyethylenimine (PEI) are used to demonstrate optical barcoding of various biological cells.

**Results**

We used two different epitaxial structures grown on 3-inch GaAs substrates. Figure 1a shows the first design, in which the active gain material is InGaP with a thickness of 178 nm and is sandwiched by 5-nm thick InGaAlP clad layers to reduce surface recombination on the outer surfaces. Additional 5-nm-thick InGaP layers were added to make the outer surface of disks free of aluminum to avoid oxidation-induced degradation in moist or aqueous environment. The second epitaxial wafer design is described later. Figure 1b shows a representative 3-inch epitaxial wafer scalably patterned with the device structures using stepper UV lithography and $O_2$ plasma ashing. Figure 1c illustrates the fabrication process by a series of scanning electron microscopy (SEM) images. The first step involves stepper UV lithography to obtain 800 nm diameter circular SU8 photoresist patterns (Fig. 1c-i). With typically available production-grade lithography (Auto Step 200, 5× reduction, i-line: 365 nm), it is generally difficult to reliably produce circular patterns below 750 nm in diameter. To solve this problem, we used oxygen ($O_2$) plasma ashing to reduce the photoresist structure to smaller dimensions (Fig. 1c-ii). The $O_2$ ashing process transforms the SU8 photoresist to gaseous carbon oxides, thereby reducing the diameter (and thickness). For a specific plasma power and $O_2$ pressure/flow rate, the final diameter of the resist is determined by the ashing time. The resist patterns are then used as etch masks for reactive ion etching (RIE) to obtain the mesa structure (Fig. 1c-iii). Finally, the resist is removed, and GaAs sacrificial layers are selectively etched using dilute piranha solution ($H_2SO_4:H_2O_2:H_2O = 1:1:100$). Partial etching results in a disk-on-pillar array on the wafer (Fig. 1c-iv). Longer etching detaches the disks completely off the wafer for producing LPs. The smallest possible diameter that can be achieved by this method, is limited by the thickness reduction of the photoresist through the ashing process, which doesn't withstand the full RIE etch time once the ashed photoresist thickness is below ~150 nm. Starting with 600 nm thick and 800 nm diameter resist, we were able to obtain disk diameters down to ~ 350 nm. Further details of the fabrication process are given in Methods and Supplementary Figs. S1-S3.

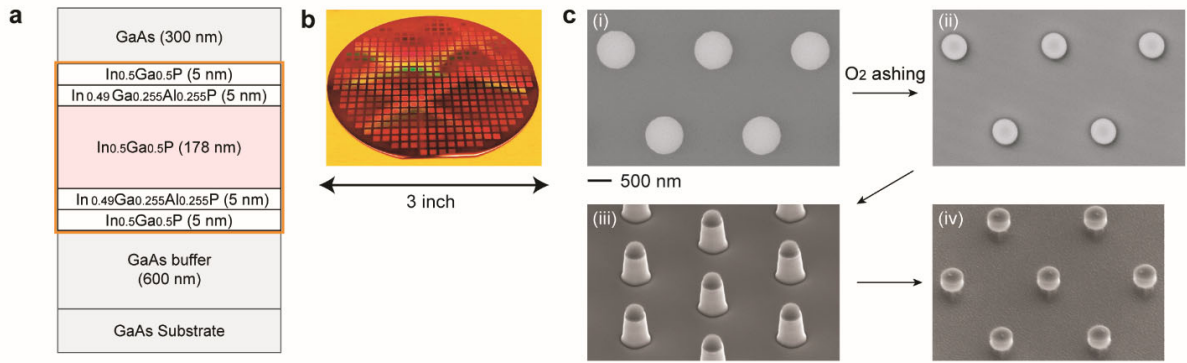

**Fig. 1 | Wafer-scale fabrication of semiconductor nanodisks**. **a** Schematic of the first wafer design. The core/clad layers that constitute laser disks are marked with the orange-line box. **b** Photo of a 3-inch wafer after stepper UV lithography and $O_2$ plasma ashing. **c** SEM images after four major fabrication steps: (i) SU8 patterns after stepper lithography, (ii) SU8 patterns with reduced diameters (to 500 nm) by $O_2$ plasma ashing, (iii) mesa structure after reactive ion etching (still capped with SU8), (iv) disk-on-pillar structure after SU8 removal and sacrificial layer partially etched.

Figure 2a shows a 460 nm diameter InGaP disk on a ~120 nm diameter GaAs pillar. We used a nanosecond optical parametric oscillator (OPO) tuned to 560 nm and a hyperspectral microscope to measure the lasing properties (Supplementary Fig. S4). Figures 2b-e show the lasing characteristics of a representative 460 nm diameter InGaP disk on GaAs pillar. A sharp single-mode lasing emission appeared at 660 nm from the disk above a threshold pump fluence of ~1 mJ/cm$^2$ (Fig. 2b). The linewidth decreased from 20 nm below threshold to 0.15-0.2 nm after lasing (Fig. 2c). The light-in-light-out plot indicates a spontaneous emission factor ($\beta$) of 0.01 (Fig. 2d). The laser far-field emission showed interference patterns (Fig. 2e). Finite difference time domain (FDTD) simulation shows that the lasing mode is of the 4-th order (m=4) whispering gallery mode (WGM) (Fig. 2f). Quality factor (Q) of the simulated mode was ~430 although the actual value may be somewhat lower.[22] We obtained similar lasing thresholds and linewidths from slightly larger disks, but no lasing was observed from disks smaller than 460 nm either on pillars or when fully detached and placed on glass substrates.

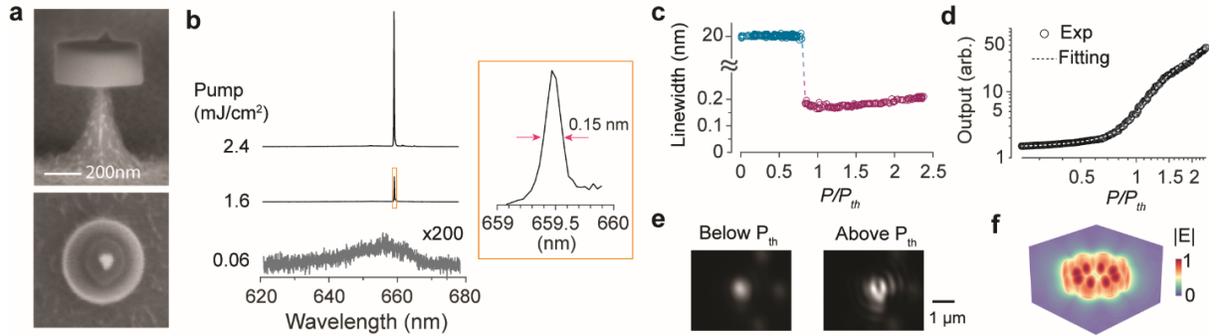

**Fig. 2 | Lasing characteristics of a 460-nm diameter InGaP disk on thin pillar. a** SEM images – side view (upper panel) and top view (lower panel) – of the InGaP disk on pillar nanolaser. **b** Emission spectra at different pump fluences. Inset, zoom-in of the spectrum at pump fluence of 1.6 mJ/cm$^2$ (threshold fluence ~1 mJ/cm$^2$). **c** Measured linewidth at different pump fluences. **d** Light-in-light-out curve. **e** Output far-field profiles. **f** FDTD simulated electric field amplitude, |E|, of the WGM mode at ~660 nm.

We tested nanodisk LPs placed on a gold substrate (Fig. 3a) with the expectation that coupling with the gold's surface plasmon polariton may allow lasing of smaller disks[20,21]. We have previously shown plasmonic lasing from submicron CsPbBr$_3$ perovskite crystals on a gold substrate with lasing thresholds lower than the same gain crystals on glass.[21] The Purcell enhancement of the plasmonic mode due to its small mode volume increased the gain and spontaneous emission factor enough to overcome the metallic loss.[20] Our numerical and FDTD simulations predicted a similar effect for InGaP disks. For example, the dielectric mode of a 360 nm diameter disk in air has a $Q$ factor of only 20 and a moderate Purcell factor of 2.3. By contrast, the same disk on gold has a much higher-order WGM plasmonic mode localized at the semiconductor-gold interface (Fig. 3b). Its $Q$ factor of 30 and Purcell factor of 19 are higher than those of the photonic mode without gold. Supplementary Note S1 mentions the laser rate equations and parameter definitions used for this analysis. To verify the Purcell enhancement experimentally, we performed time-resolved photoluminescence (PL) spectroscopy. PL decay curves were obtained from three different samples: intact wafer, a 380 nm-sized disk on pillar, and a disk of a similar size on gold substrate (Fig. 3c). The bulk semiconductor from the full wafer stack exhibited a monotonous exponential decay with a lifetime of 1840 ps. The disk on pillar showed a mono-exponential decay but with a shorter lifetime of 623 ps, an evidence of cavity-enhanced Purcell effect. The disk-on-gold sample showed two distinct exponential decays. The first fast decay has a lifetime less than the photodiode's response time (<35 ps) and is attributed to plasmonic modes strongly coupled with the gold's plasmons. The second slow decay with a lifetime of 410 ps is thought to be associated with other photonic-like modes weakly interacting with the plasmons. We observed lasing from 360-nm diameter disks on gold, but not on pillar or glass. Figures 3d and 3e show the output spectra of two representative plasmonic devices. The threshold pump fluences were 0.9 and 1.5 mJ/cm$^2$ respectively, similar to the smallest (460 nm diameter) lasers on pillar. The linewidth above threshold was about 1.5 nm, considerably broader than the all-dielectric lasers but narrower than PL linewidth below threshold. Figure 3f shows the far-field emission profiles of a plasmonic nanolaser below and above lasing threshold.

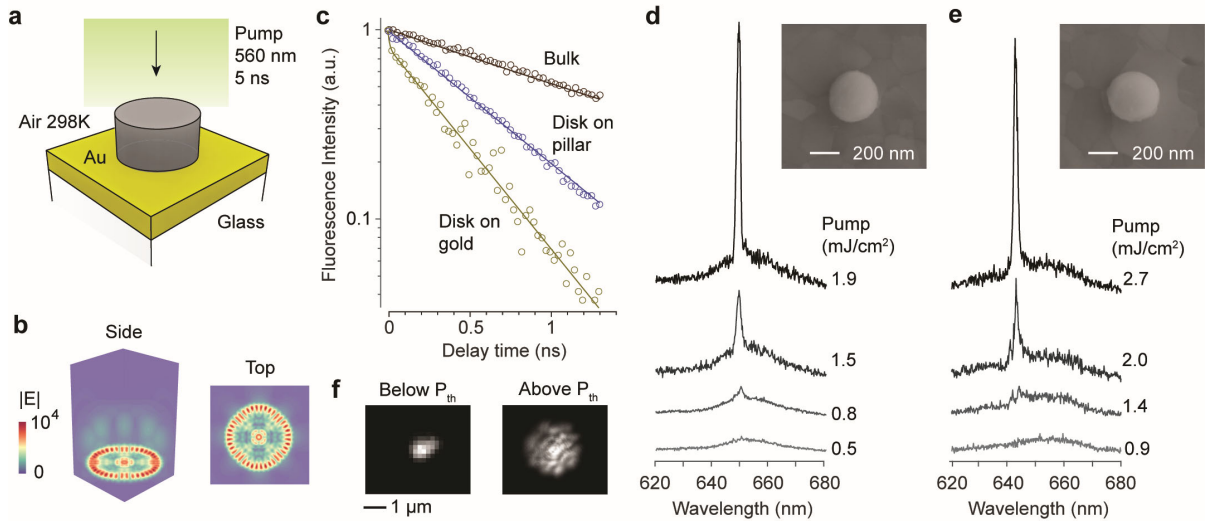

**Fig. 3 | Surface plasmon polariton enhanced nanolasers. a** Schematic of a disk-on-gold laser. **b** FDTD simulated electric field profile, |E|, of the plasmonic mode for a 360-nm-diameter InGaP disk on gold. **c** Time-resolved PL decay curves of the bulk material in the wafer (brown), a disk on pillar (blue) and a disk on gold (yellow). Lines, double-exponential fits. **d** and **e** Measured output spectra of two representative disks on gold. Insets: top-view SEM images of the nanolasers. **f** Measured far-field emission profiles below and above lasing threshold.

To produce LPs over a wide spectral range, we used a multi-disk wafer design[23] consisting of two disk regions separated by GaAs sacrificial layers (Fig. 4a). The first disk region is identical to the single-InGaP-active-layer wafer described earlier except for a slightly thinner gain region (158 nm). The second disk region is comprised of an InGaAsP core and InGaP clad layers. The InGaAsP-core gain region layer has a PL peak at 706 nm, ~56 nm red-shifted from the InGaP-core gain region layer. Using the process described earlier, mesa structures with diameters in the range of 500 to 700 nm were produced. After removing the resist, the entire chip was immersed in dilute piranha solution to selectively etch the GaAs sacrificial layers. Following this, through a multi-step centrifugation-resuspension process, we finally obtained a suspension of the nanodisks in ethanol. Finally, they were fully coated with a silica shell of ~25 nm thickness. Figure 4b shows the SEM image of a silica-coated LP with a diameter of ~ 650 nm semiconductor core and ~ 25 nm silica shell. The theoretical wavelength tuning slopes as a function of diameter are 0.81, 0.92, and 1.04 nm/nm for the WGM modes with m = 4, 5, and 6, respectively (Fig. 4c). The intended variation of the disk diameters resulted in LPs with a continuum of emission peaks spanning over an 80 nm range from 635 to 715 nm. Figure 4d shows a collection of laser spectra from different LPs suspended in water. Figures 4e and 4f show the wavelength histograms of LPs from the two individual disk regions, obtained by controlled wet etching to harvest LPs from each disk region separately.

The silica-coated LPs were tested for cell tagging after surface functionalization with polyethylenimine (PEI) for efficient uptake.[11] Figure 4g shows RAW 264.7 macrophages, MCF-7 breast cancer cells, and Jurkat T cells in culture, after overnight incubation with LPs. Since the uptake process is stochastic, each cell can internalize a different number of LPs. The presence of LPs in the cytoplasm has no statistically significant effect on the viability of the cells over 72 hours (Supplementary Fig. S5), and the cells continue to proliferate and thrive with natural cadence. Figure 4h demonstrates lasing of these LPs from within the cells. The set of LPs associated with each cell acts as their optical barcode, which can be used to uniquely identify individual cells and track them spatiotemporally for large-scale single-cell analysis.[10,11]

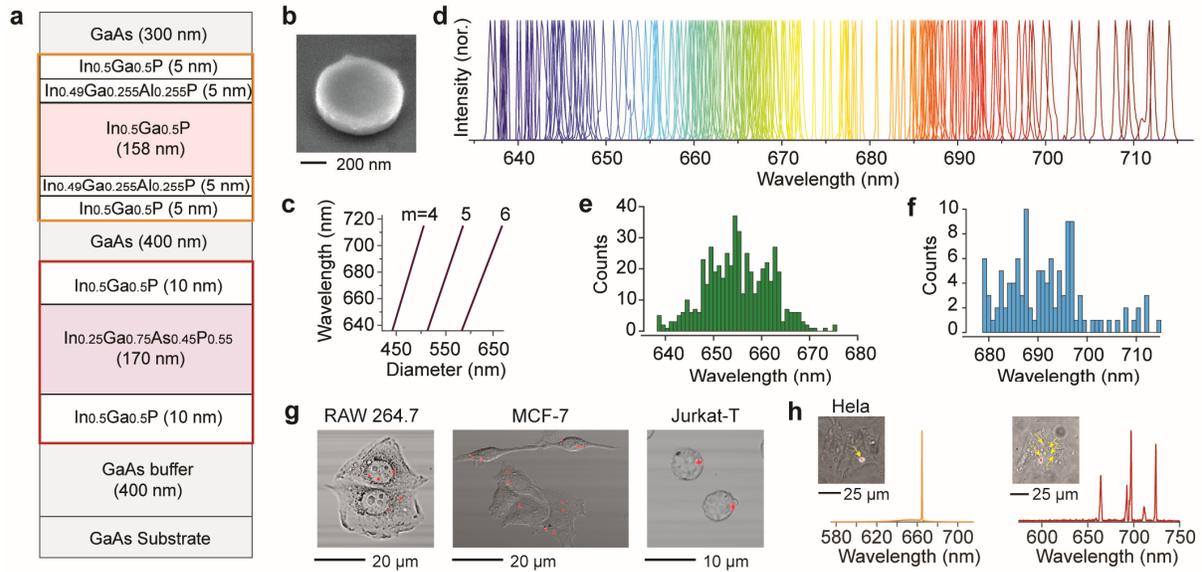

**Fig. 4 | Multi-color nanodisk particles and intracellular lasing. a** Schematic of the second wafer. The two disk regions are marked with orange and red boxes, respectively. **b** SEM of a silica-coated nanodisk particle. **c** FDTD-calculated resonance wavelengths of WGM modes with mode order m for different disk diameters. **d** Normalized laser emission spectra of representative ~100 LPs. **(e-f)** Histogram of lasing wavelengths of LPs from **e** top and **f** bottom layers. **g** Overlaid bright-field and confocal fluorescence image (red) of LPs inside various biological cells in culture. **h** Emission spectra and overlaid bright-field and stimulated emission images of single or 5 LPs inside living HeLa cells.

**Discussion**

Figure 5 compares our results with various previous nanolasers in both, dielectric (semiconductor and no metal) and plasmonic (semiconductor and metal) designs. The bulk-core and clad structure of our semiconductor nanolasers provided higher optical gain than quantum-well structures, which allowed lasing of InGaP disks as small as 460 nm in diameter. To the best of our knowledge, this represents the smallest III-V semiconductor laser demonstrated to date and the smallest WGM laser made of any materials. Furthermore, the plasmonic lasing achieved with InGaP disks on gold with 360-nm diameter and 198-nm height (~0.02 $\mu m^3$ in volume) is, in terms of semiconductor volume, the smallest plasmonic laser ever demonstrated at room temperature and is comparable to other nanolasers operated at cryogenic temperatures (Supplementary Note. S2). It should also be noted that this work represents the first optical lithography-based fabrication of submicron-sized LPs. Our results show the potential of multi-color nanodisk LPs in the silicon-detector range for large-scale optical barcoding of cells.

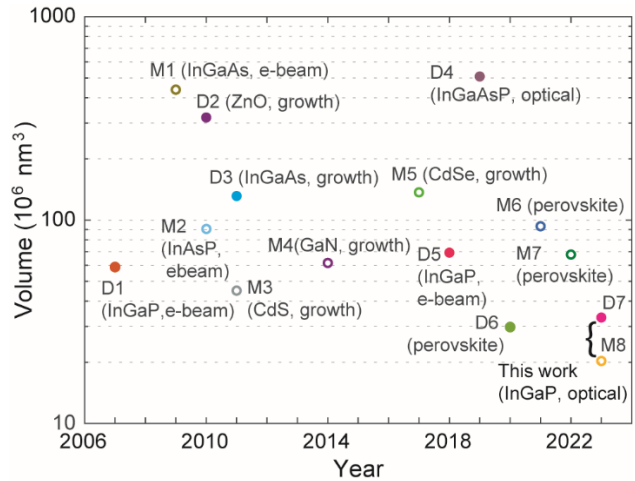

**Fig. 5 | Device volume of semiconductor-material based various dielectric and plasmonic nanolasers operating at room temperature.** Solid circles represent semiconductor-only dielectric lasers, and hollow circles represent metal-incorporated plasmonic lasers. Each data point is labeled with the device name used in Supplementary Table S1 and S2, and the gain medium and fabrication method (in parentheses).

**Methods**

**Epitaxial wafer structure.** The devices are fabricated from two different wafer structures. In both cases, the substrate is GaAs. In the first wafer structure, only a single active layer material is present, with the exact structure being: GaAs substrate / 600 nm GaAs/ 5 nm InGaP /5 nm $In_{0.49}(Ga_{0.5}Al_{0.5})_{0.51}P$ /178 nm InGaP /5 nm $In_{0.49}(Ga_{0.5}Al_{0.5})_{0.51}P$ /5 nm InGaP /300 nm GaAs. The central InGaP forms the main active material layer while the 10 nm layers on either side form a cladding layer to reduce surface recombination. The top 300 nm GaAs is a sacrificial capping layer to protect the active material from potential damage during fabrication. In the second wafer structure, two active layer materials are present, with the exact structure being: GaAs substrate / 400 nm GaAs / 10 nm InGaP/ 170 nm InGaAsP/ 10 nm InGaP/ 400 nm GaAs/ 5 nm InGaP/ 5 nm $In_{0.49}(Ga_{0.5}Al_{0.5})_{0.51}P$/ 158 nm InGaP/ 5 nm $In_{0.49}(Ga_{0.5}Al_{0.5})_{0.51}P$/ 5 nm InGaP/ 300 nm GaAs. GaAs layers are sacrificial. These are custom ordered and epitaxially grown using metal organic chemical vapor deposition (MOCVD) by Xiamen Powerway Advanced Materials Co. Ltd.

**Fabrication of nanodisk lasers on chip.** Device fabrication starts with cleaning the wafers in acetone, isopropyl alcohol (IPA), methanol, and deionized water to remove dissolvable contaminants on the surface. The wafer is baked at 180 °C to dehydrate adsorbed water molecules and then oxygen plasma processed to descum surface organics and to improve photoresist adhesion. SU8 2000.5 is then spin-coated and soft-baked to form a 500 nm thick layer. Exposure is performed using stepper UV lithography (Auto Step 200) followed by post-exposure bake and development using propylene glycol methyl ether acetate (PGMEA) and rinsed with IPA. SU8, being a negative photoresist, retains the patterns at the locations where it is exposed. The patterns are then hard-baked to 180 °C and then the wafer is descummed using oxygen plasma to remove any remaining photoresist molecules on the unexposed substrate surface. To get sub-micron dimensions, these photoresist patterns are controllably ashed using oxygen plasma (Anatech SP100) where

~ 1 min ashing leads to a reduction in diameter by ~100 nm. Reactive ion etching (RIE) using a chlorine-based chemistry is then carried out in SAMCO 200iP or 230 iP to create III-V micropillars with the patterned hard-baked photoresist acting as etch mask. A methane-chlorine-based chemistry used in an Oxford RIE tool gave significant anisotropy as shown in Supplementary Fig. S6a, compared to the morphology obtained using the boron trichloride-argon chemistry in SAMCO RIE represented by the SEM image in Supplementary Fig. S6b. The photoresist mask is then removed using a carbon tetrafluoride-oxygen plasma followed by dissolution in sulfuric acid solution. Dilute piranha solution ($H_2SO_4:H_2O_2:H_2O::1:1:100$) is used to selectively etch the GaAs layers while keeping the active materials intact. Partial etching of the GaAs layers creates a pillar structure with the diameter smaller than the overlying disk, thereby creating the nanodisk structures on chip. Supplementary Fig. S7 shows a representative pillar structure fabricated from the bilayer wafer, with the GaAs layers partially etched.

**Fabrication of free-standing laser particles.** After creating the nanodisk structures on chip, complete etching of GaAs releases the nanodisks in the piranha solution. The next steps involve collection of the laser particles and making them biocompatible for cell studies. The laser particle suspension in piranha is centrifuged to collect the disks at the bottom of a tube and the supernatant liquid is selectively removed. Ethanol is then added to the tube and the laser particles are redispersed in it by ultrasonication. Repeating this centrifugation, decanting of supernatant, adding fresh ethanol and resuspension of laser particles, the medium is effectively cleared of piranha solution. Using tetraethyl orthosilicate (TEOS) with the reaction initiated in a basic medium (through ammonium hydroxide), a uniform silica shell is then created around each disk to obtain freely suspended biocompatible silica-coated laser particles. These are then surface-functionalized with polyethylenimine (PEI) for improved cellular association.

**Optical characterization.** For laser experiment, the specimen is placed in a home-built epi-fluorescence microscopy setup. The pump source is an optical parametric oscillator (OPO, Optotek HE 355 LD) tuned to 560 nm, with a repetition rate of 20 kHz and a pulse duration of 4 ns. Using a 0.6 NA, 50× air objective lens or 0.9 NA, 100× air objective lens (Nikon), the full-width-at-half-maxima (FWHM) size of the pump beam on the sample is about 20 μm. The emission from the sample collected by the objective lens is passed through a dichroic mirror and a dichroic filter and split to a silicon-based EMCCD camera (Luca, Andor) for wide-field imaging and to the grating-based EMCCD spectrometer (Shamrock, Andor). With an entrance slit width of 20 μm, the measurement spectral resolution is about 0.13 nm. For time-resolved photoluminescence measurements, a picosecond laser (VisIR-765, PicoQuant) is used, which is frequency doubled to 382 nm using a nonlinear Beta barium borate (BBO) crystal, a single-photon avalanche photodiode (MicroPhotonics Devices) with a single photon timing resolution (FWHM) of 35 ps, and a time-correlated single-photon counting board (TimeHarp 260, PicoQuant).

**Fabrication and characterization of plasmonic laser.** The free-standing disks suspended in ethanol solution is simply drop-cast to a polycrystalline gold substrate (Platypus). The root-mean-square roughness of the gold surface is about 0.7 nm.[21] After characterizing the lasing properties of the device, we marked the corresponding device's position by irradiating the gold substrate using high pump fluence (Supplementary Fig. S8). The devices were then imaged with SEM to precisely quantify their sizes.

**Finite difference time domain (FDTD) simulation.** The eigenmodes and field profiles of the nanodisk lasers are numerically simulated with a commercial 3D finite-difference time-domain (3D FDTD) solver on Lumerical. The optical constants for gold are taken from CRC handbook.[24]

**Cell experiment.** HeLa human cervical cancer cells, MCF-7 human breast cancer cells, RAW 264.7 mouse macrophage cells, and Jurkat, Clone E6-1 human T lymphoblast (Jurkat T) cells are purchased from ATCC (American Type Culture Collection). HeLa, MCF-7, and RAW 264.7 cells are cultured and maintained in the cell media of Dulbecco's modified Eagle medium (DMEM) supplemented with 10% (v/v) fetal bovine serum (FBS) and 1% (v/v) penicillin-streptomycin at 37 °C under 5% $CO_2$. Jurkat T cells are cultured in RPMI 1640 Medium (Thermo Fisher) cell media supplemented with 10% FBS and 1% antibiotic-antimycotic. For cytotoxicity assessment, HeLa and RAW 264.7 cells are seeded in 96-well plastic plates with the density of 3000 cells/well and incubated first for 24 hours. Then about 6,000 silica-coated visible LPs are added into each well. After further incubation for 24 h, 48 h, and 72 h, the standard cell counting kit (CCK8) (Sigma-Aldrich) assay is performed to test the cell viability. The non-treated cells at different time points are used as control (100%) for calculation. For each group, six parallel experiments are performed simultaneously. For the cell tagging experiment, cells are plated in their respective media at a known density in a glass-bottom plate. After 24 hours of incubation, the visible laser particles resuspended in PBS buffer are added into cells with the particle-to-cell ratio 4:1. The plate is shaken to mix the media evenly and the cell media is carefully refreshed 6 hours later. After 24 hours of further incubation, the cells are washed three times with DPBS buffer and fixed using a 4% Paraformaldehyde (PFA) solution for following confocal imaging (Olympus, FV3000) and cell laser experiment.

## Data availability

The data that support the findings of this study are available from the corresponding author upon reasonable request.

## Acknowledgements

This study was supported by the US National Institutes of Health research grants (DP1-OD022296, R01-EB033155). SC acknowledges the MGH fund for medical discovery (FMD) fundamental research fellowship award. This research used the resources of the Center for Nanoscale Systems, part of Harvard University, a member of the National Nanotechnology Coordinated Infrastructure (NNCI), which is supported by the National Science Foundation under award number 1541959, and of the MIT.nano facilities, part of Massachusetts Institute of Technology.

## Author contributions

DS and SC contributed equally to this work. DS, SC, and SHY designed the project. DS developed and performed nanofabrication. SC performed optical measurements and computer simulation. HY conducted cell culture. NM and PH contributed to nanofabrication and measurement. DS, SC, and SHY wrote the manuscript with input from all authors.

## Competing interests

PHD, NM, and SHY hold patents on laser particle technologies. NM and SHY have financial interests in LASE Innovation Inc., a company focused on commercializing technologies based on laser particles. The financial interests of NM and SHY were reviewed and are managed by Mass General Brigam in accordance with their conflict-of-interest policies.

# Supplementary Information for

# Ultrasmall InGa(As)P dielectric and plasmonic nanolasers

Debarghya Sarkar[1,2†], Sangyeon Cho[1,2†], Hao Yan[1,2], Nicola Martino[1,2], Paul H. Dannenberg[1-3], Seok-Hyun Yun[1-3*]

†Equal Contribution
*Corresponding Author: syun@hms.harvard.edu

[1]Harvard Medical School, Boston, MA 02115, USA
[2]Wellman Center for Photomedicine, Massachusetts General Hospital, Boston, MA 02114, USA
[3]Harvard-MIT Health Sciences and Technology, Massachusetts Institute of Technology, Cambridge, MA 02139, USA

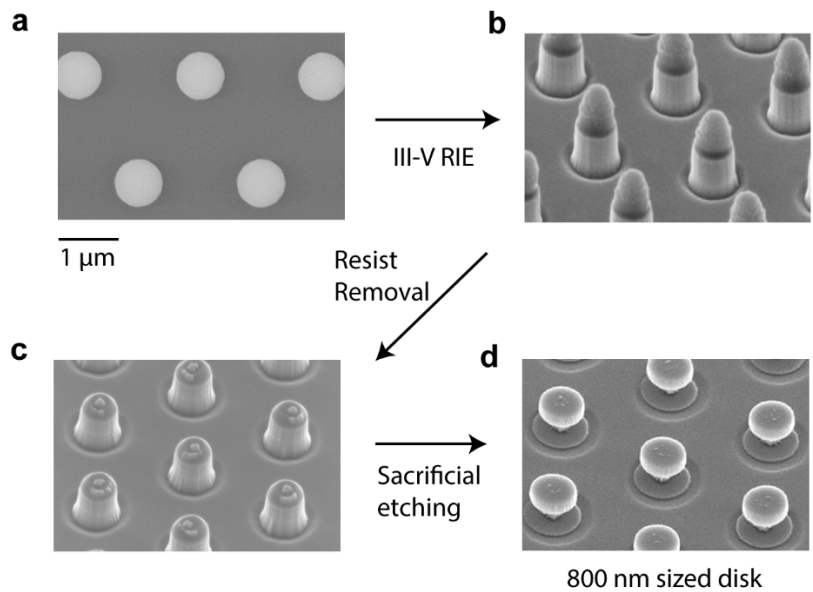

**Supplementary Figure S1**. **Fabrication of 800 nm diameter disk**. SEM images of fabrication of 800 nm diameter disks **a** SU8 patterns using stepper lithography, **b** micropillars with resist mask on top after RIE, **c** micropillars with resist mask removed, and **d** microdisks with resist removed and sacrificial layer partially etched to create disk-on-pillar structures.

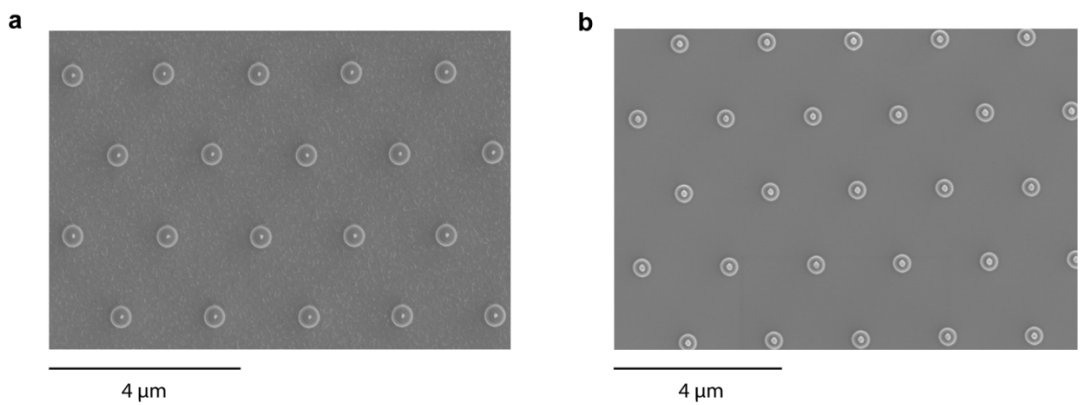

**Supplementary Figure S2. Top view SEM image of disks-on-pillar.** Top view SEM images of **a** 450 nm, and **b** 400 nm diameter disks-on-pillar. The small dot on top of each disk is the partially etched GaAs

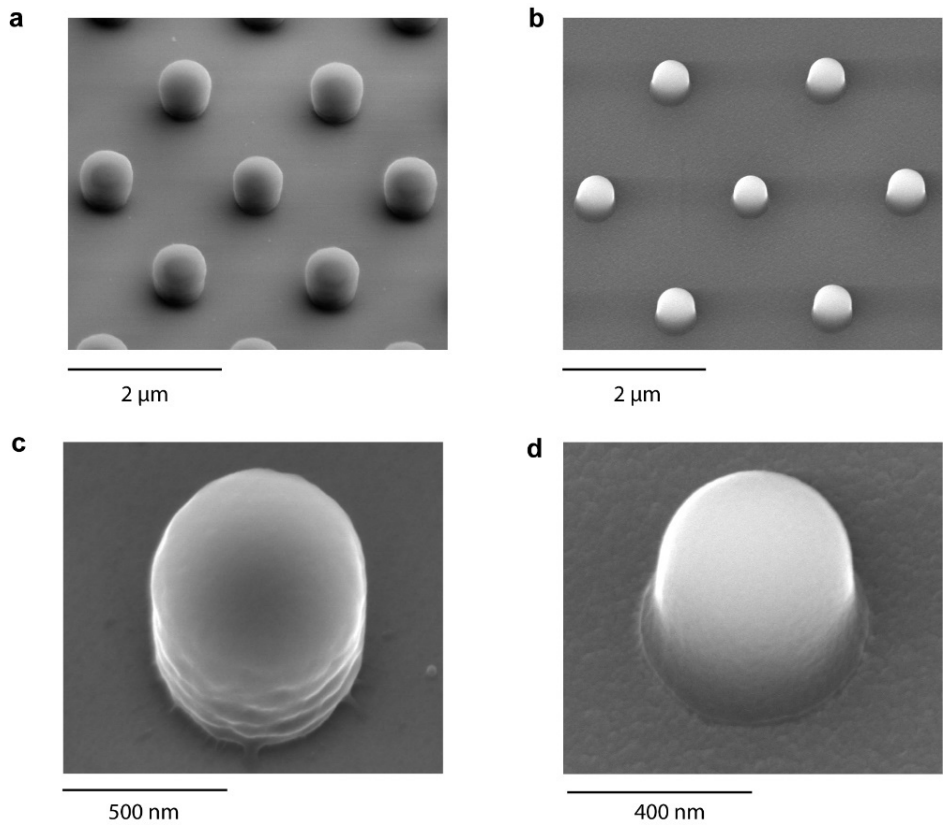

**Supplementary Figure S3. Resist geometry control by $O_2$ plasma ashing**. **a** Tilt view of stepper lithography ~800 nm diameter resist patterns. **b** Tilt view of patterns after hard baking and $O_2$ plasma ashing. **c** Single resist pillar before hard baking showing interference patterns on side wall from UV exposure. **d** Single resist pillar after hard baking and $O_2$ plasma ashing showing smooth sidewall.

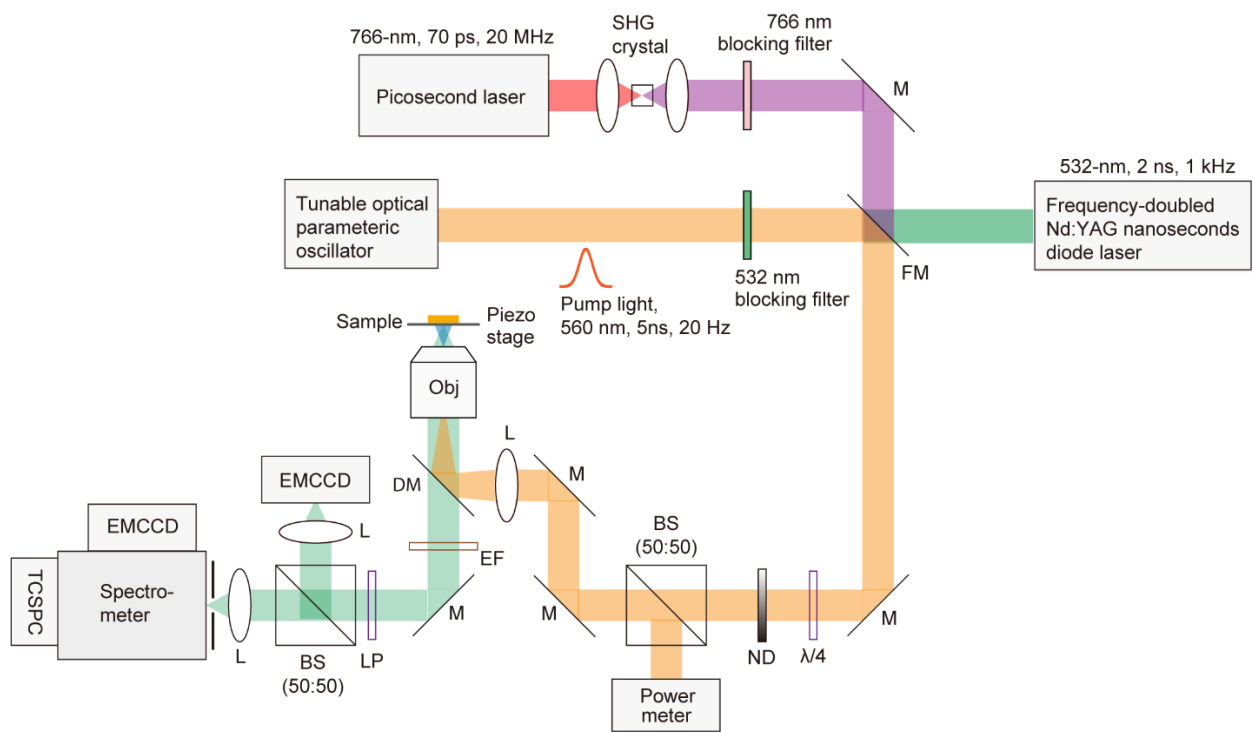

**Supplementary Figure S4. Laser optical characterization setup**

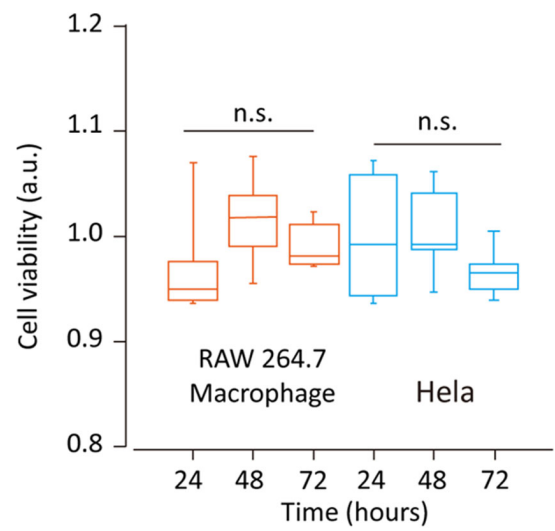

**Supplementary Figure S5.** Cell viability of RAW 264.7 macrophage and HeLa cells.

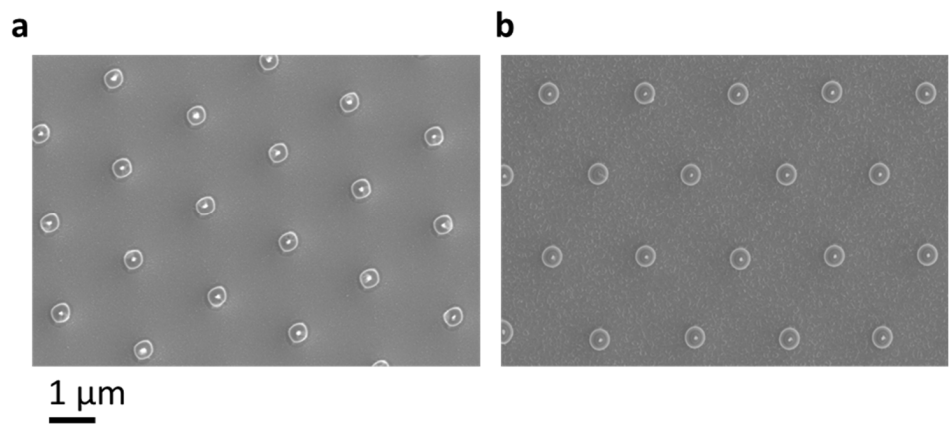

**Supplementary Figure S6. Different reactive ion etching processes. a,** Morphology of disks using $CH_4$-$H_2$-$Cl_2$ etch in Oxford RIE; **b,** Morphology of disks using $BCl_3$-Ar etch in SAMCO RIE.

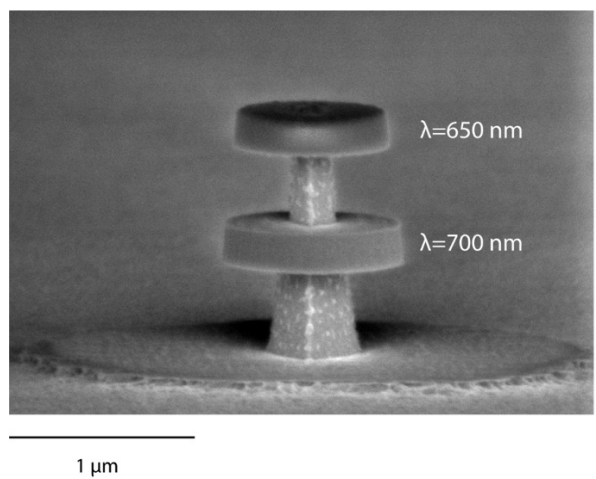

**Supplementary Figure S7.** Bilayer disk structure on wafer with sacrificial layers partially etched.

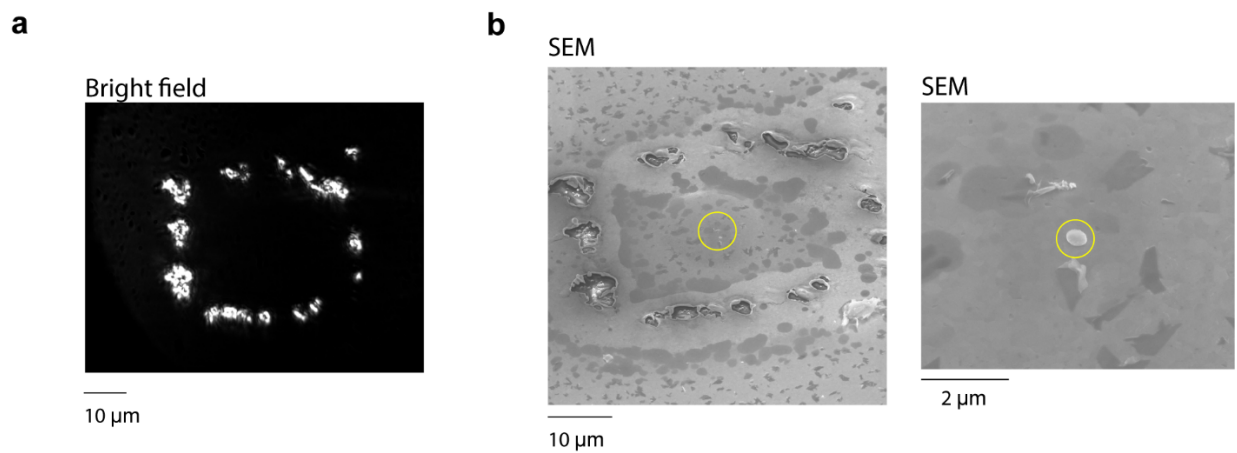

**Supplementary Figure S8. Marking samples for SEM. a** Bright-field optical image of a gold substrate after generating patterns on the substrate using high power pump beam to mark the location of a plasmonic disk on the substrate. **b** SEM images of the corresponding sample (circle) identified by the particular pattern on the substrate.

**Supplementary Note S1. Disk laser rate equations and definitions of parameters**

The rate equation for the semiconductor laser is given by

$$\frac{d}{dt}\rho(t) = P(t) - \frac{\beta VF}{\tau_s}[\rho(t) - \rho_{tr}]q(t) - \frac{F}{\tau_s}\rho(t) - \frac{1}{\tau_{nr}}\rho(t) \qquad (1)$$

$$\frac{d}{dt}q(t) = \frac{\beta VF}{\tau_s}[\rho(t) - \rho_{tr}]q(t) + \frac{\beta F}{\tau_s}\rho(t) - \frac{1}{\tau_c}q(t) \qquad (2)$$

Where $P(t)$ is the rate of pumping, $\rho(t)$ is the carrier density, $\rho_{tr}$ is transparent carrier density (about $10^{18}$ cm$^{-3}$), $q(t)$ is the number density of the cavity photon, $\beta$ is the spontaneous emission factor, $\tau_s$ and $\tau_{nr}$ are the radiative and the non-radiative lifetime of the carrier, respectively, $\tau_c$ is the photon lifetime inside the cavity. Here, the Purcell factor $F$ is defined as

$$F = \frac{3}{4\pi^2}\left(\frac{Q}{V_m}\right)\left(\frac{\lambda}{n}\right)^3 \qquad (3)$$

where $Q$ is the quality factor, $\lambda$ is the resonance wavelength, $n$ is the refractive index, $V_m$ is the mode volume described by $W(r)$, the electromagnetic energy density of the mode[1]

$$V_m = \frac{\int_{gain} W(r)d^3r}{\max\,[W(r)_{gain}]} \qquad (4)$$

$$W(r) = \frac{1}{4}\left\{Re\left[\frac{d(\omega\varepsilon)}{d\omega}\right]|E(r)|^2 + \mu|H(r)|^2\right\} \qquad (5)$$

The steady-state solution about the unity quantum yield gain can be expressed as

$$P_{th}V = \frac{\omega_0}{\beta Q} + \frac{F\rho_{tr}V}{\tau_s}(1-\beta) \qquad (6)$$

The first term describes the radiative cavity loss, and the second term is related about the population inversion. Using the fact that the threshold condition is equivalent to having one photon in the cavity mode volume ($qV = 1$), the number of the threshold carrier $\rho_{th}V$ is estimated as

$$\rho_{th}V = \frac{\tau_s\omega_0}{(1+\beta)\beta FVQ} + \frac{\rho_{tr}V}{1+\beta} \approx \frac{\tau_s\omega_0}{\Gamma\beta FQ} + \rho_{tr}V \qquad (7)$$

Despite the low $Q$, the plasmonic devices could have lower threshold carrier density by enhanced optical gain by large β factor and Purcell effect[2,3].

**Supplementary Note S2. Comparison with other representative all-dielectric and plasmonic micro- and nanolasers operating at room-temperature and cryogenic temperature**

A list of this work and the previously demonstrated room-temperature dielectric (Supplementary Table S1) and plasmonic lasers (Supplementary Table S2) and cryogenic nanolasers (Supplementary Table S3) to date[4,5].

**Supplementary Table S1. A list of representative room-temperature dielectric nanolasers to date.**

| | Cavity Structure (type) | Gain material (Wavelength) | Gain-medium size | Lasing modality | Fabrication Method | Threshold (Reported value, method) | Pulse duration (Repetition) | Year (Ref) |
|---|---|---|---|---|---|---|---|---|
| D1 | Disk (on pillar) | InGaP quantum well with InGaAlP cladding (660 nm) | 645 nm (diameter) x 180 nm (height) | WGM | E-beam lithography | (70 mJ/cm$^2$, Optical) | 8 ns (33 kHz) | 2007 ([6]) |
| D2 | Hexagonal nanodisk | ZnO on silica (390 nm) | 490 nm (side) x 550 nm (height) | WGM | Chemical vapor transport growth | (2 mJ/cm$^2$, Optical) | 5 ns (10 Hz) | 2010 ([7]) |
| D3 | Hexagonal nanopillar | InGaAs core in GaAs shell (900 nm) | 270 nm (diameter) x 2 μm (height) | Helical | Nanowire growth | (22 μJ/cm$^2$, Optical) | 120 fs (76 MHz) | 2011 ([8]) |
| D4 | Disk (free standing) | InGaAsP (1170~1580 nm) | 1800 nm(diameter) x 250 nm (height) | WGM | Optical lithography | (56 μJ/cm$^2$, Optical) | 3 ns (2 MHz) | 2019 ([9]) |
| D5 | Disk (free standing) | InGaP | 700 nm (diameter) x 180 nm (height) | WGM | E-beam lithography | (30 uJ/cm$^2$, Optical) | 1.5 ns (100 Hz) | 2019 ([10]) |
| D6 | Cube (on substrate) | CsPbBr$_3$ (530 nm) | 310 nm (side) | Mie | Solution deposition | (300 uJ/cm$^2$, Optical) | 150 fs (100 kHz) | 2020 ([11]) |
| **D7** | **Disk (on pillar)** | **InGaP bulk (650 nm)** | **460 nm (diameter) x 200 nm (height)** | **Whispering Gallery Mode (WGM)** | **Optical lithography** | **(1 mJ/cm$^2$, Optical)** | **5 ns (20 Hz)** | **This work** |

**Supplementary Table S2. A list of representative plasmonic lasers demonstrated to date.**

| | Cavity Structure | Gain materials (Wavelength) | Gain-medium size | Metal | Lasing modality | Fabrication Method | Threshold (Reported value, method) | Pulse duration (Repetition) | Year (Ref) |
|---|---|---|---|---|---|---|---|---|---|
| M1 | Rectangular pillar (metal-coated, on substrate) | InGaAs (1500 nm) | 310 nm (diameter) 6000 nm (length) x 300 nm (height) | Silver | Fabry-Perot | E-beam lithography | (5.9 mA, electrical) | 28 ns (1 MHz) | 2009 ([12]) |
| M2 | Disk (metal-coated, on substrate) | InAsP quantum well (1300 nm) | 700 nm (diameter) x 235 nm (height) | Silver | WGM | E-beam lithography | (120 kW/cm$^2$, Optical) | 60 ps (80 MHz) | 2010 ([13]) |
| M3 | Square Plates (on metal substrate) | CdS (500 nm) | 1 μm (side length) x 45 nm (thickness) | Silver | WGM | Vapor-phase growth | 200 μJ/cm$^2$ (2 GW/cm$^2$, optical) | 100 fs (10 kHz) | 2011 ([14]) |
| M4 | Nanowires (on metal substrate) | GaN (375 nm) | 100 nm (diameter) x 15 μm (length) | Aluminum | Fabry-Perot | Vapor-phase growth | 35 mJ/cm$^2$ (3.5 MW/cm$^2$, optical) | 10 ns (100 kHz) | 2014 ([15]) |
| M5 | Square Plates (on metal substrate) | CdSe (700 nm) | 1 μm (side length) x 137 nm (height) | Gold | WGM | Vapor-phase growth | 45 μJ/cm$^2$ (10 kW/cm$^2$, optical) | 4.5 ns (1 kHz) | 2017 ([3]) |
| M6 | Cubes (on metal substrate) | CsPbBr$_3$ (540 nm) | (Device 1. Length: 570 nm, Height: 320 nm Device 2. Length: 720 nm, Height: 180 nm) | Gold | WGM | Sonochemistry | (1. 1 mJ/cm$^2$ 2. 0.7 mJ/cm$^2$, Optical) | 5 ns (20 Hz) | 2021 ([2]) |
| M7 | Square Plates (on metal substrate) | CsPbBr$_3$ (540 nm) | 1.06 μm (length) x 0.83 μm (length) x 77 nm (height) | Silver | WGM | Solution deposition | (26 μJ/cm$^2$) | 190 fs (80 MHz) | 2022 ([13]) |
| **M8** | **Disk (on substrate)** | **InGaP bulk (650 nm)** | **360 nm (diameter) x 200 nm (height)** | **Gold** | **WGM** | **Optical lithography** | **(0.9~1.5 mJ/cm$^2$, Optical)** | **5 ns (20 Hz)** | **This work** |

**Supplementary Table S3. A list of representative nanolasers at cryogenic temperature.**

| | Cavity Structure | Gain materials (Wavelength) | Gain-medium size | Metal | Lasing modality | Fabrication Method | Temperature | Threshold (Reported value, method) | Pulse duration (Rep.) | Year (Ref) |
|---|---|---|---|---|---|---|---|---|---|---|
| D8 | Disk (on pillar) | InAs QD in GaAs (860 nm) | 627 nm (diameter) x 265 nm (height) | | WGM | Optical lithography and wet etching | 10 K | (50 uW, Optical) | 200 fs (76 MHz) | 2009 ([16]) |
| D9 | Nanocylinder | GaAs on quartz (825 nm) | 500 nm (diameter) X 330 nm (height) | | Quasi Bound States in Continuum (BIC) | E-beam lithography | 77 K | (433 μJ/cm$^2$, Optical) | 200 fs (20 kHz) | 2020 ([17]) |
| M9 | Rectangular pillar (metal-coated, on substrate) | InGaAsP (1400 nm) | 210 nm (diameter) X 300 nm (height) | Gold | Fabry-Perot | E-beam lithography | 77 K | (4 μA, electrical) | Continuous | 2007 ([18]) |
| M10 | Nanowires (on metal substrate) | CdS (489 nm) | 129 nm (diameter) x 12 μm (length) | Silver | Fabry-Perot | Vapor-phase growth | < 10K | 500 nJ/cm$^2$ (5MW/cm$^2$, optical) | 100 fs (80 MHz) | 2009 ([19]) |
| M11 | Coaxial (metal-coated, on substrate) | InGaAsP quantum well (1427nm) | 200 nm (inner core dia.) x 100 nm (gain medium ring thickness) x 200 nm (height) | Silver 98% and aluminium 2% | WGM | E-beam lithography | 4.5 K | Thresholdless | Continuous | 2012 ([20]) |